%% file: main.tex
\documentclass[sigconf,screen]{acmart}

\AtBeginDocument{%
  }

\setcopyright{acmlicensed}
\copyrightyear{2018}
\acmYear{2018}
\acmDOI{XXXXXXX.XXXXXXX}

\acmConference[Conference acronym 'XX]{Make sure to enter the correct
  conference title from your rights confirmation emai}{June 03--05,
  2018}{Woodstock, NY}

\acmISBN{978-1-4503-XXXX-X/18/06}

\usepackage{booktabs}
\usepackage{multirow}
\usepackage{tabularx}
\newcolumntype{Y}{>{\raggedright\arraybackslash}X}
\newcolumntype{M}[1]{>{\centering\arraybackslash}m{#1}}
\usepackage{graphicx}
\usepackage{amsmath}
\usepackage{array}
\usepackage{pgfplots}
\pgfplotsset{compat=1.18}
\usepackage{pgfplotstable}
\usepackage{tikz}
\usetikzlibrary{calc,arrows.meta}

\definecolor{nodefill}{HTML}{2D8CFF}   
\definecolor{nodeborder}{HTML}{0F3D91} 
\definecolor{posedge}{HTML}{5F6B7A}    
\definecolor{negedge}{HTML}{C0392B}    
\definecolor{labelbg}{HTML}{FFFFFF}    

\tikzset{
  corrnode/.style={
    circle, draw=nodeborder, very thick,
    fill=nodefill, minimum size=12mm,
    inner sep=0pt, text=white, font=\bfseries
  },
  corrpos/.style={draw=posedge, line cap=round},
  corrneg/.style={draw=negedge, line cap=round}
}

\newcommand{\corrEdge}[5]{%
  \pgfmathsetmacro{\absr}{abs(#4)}%
  \pgfmathsetmacro{\lw}{0.6 + ((\absr-0.3)/(1.0-0.3))*(3.0-0.6)}%
  \draw[#5, line width=\lw pt]
    (#1) -- node[midway, sloped, fill=labelbg, inner sep=1pt, font=\footnotesize]{#3} (#2);
}

\usepackage{xcolor}
\usepackage{threeparttable}
\usetikzlibrary{calc}
\begin{document}

\title{CollaClassroom: An AI-Augmented Collaborative Learning Platform with LLM Support in the Context of Bangladeshi University Students}

\author{Salman Sayeed}
\email{salmansayeed5345@gmail.com}
\affiliation{%
  \institution{Department of CSE, Bangladesh University of Engineering and Technology}
  \city{Dhaka}
  \country{Bangladesh}
}

\author{Bijoy Ahmed Saiem}
\email{bijoysaeem@gmail.com}
\affiliation{%
  \institution{Department of CSE, Bangladesh University of Engineering and Technology}
  \city{Dhaka}
  \country{Bangladesh}
}

\author{Al-Amin Sany}
\email{a.a.sany105@gmail.com}
\affiliation{%
  \institution{Department of CSE, Bangladesh University of Engineering and Technology}
  \city{Dhaka}
  \country{Bangladesh}
}

\author{Sadia Sharmin}
\email{sadia@teacher.cse.buet.ac.bd}
\affiliation{%
  \institution{Department of CSE, Bangladesh University of Engineering and Technology}
  \city{Dhaka}
  \country{Bangladesh}
}

\author{A.~B.~M.~Alim Al Islam}
\email{alim_razi@cse.buet.ac.bd}
\affiliation{%
  \institution{Department of CSE, Bangladesh University of Engineering and Technology}
  \city{Dhaka}
  \country{Bangladesh}
}

\begin{abstract}
CollaClassroom is an AI-enhanced platform that embeds large language models (LLMs) into both individual and group study panels to support real-time collaboration. We evaluate CollaClassroom with Bangladeshi university students (N = 12) through a small-group study session and a pre–post survey. Participants have substantial prior experience with collaborative learning and LLMs and express strong receptivity to LLM-assisted study (92\% agree/strongly agree). Usability ratings are positive, including high learnability (67\% “easy”), strong reliability (83\% “reliable”), and low frustration (83\% “not at all”). Correlational analyses show that participants who perceive the LLM as supporting equal participation also view it as a meaningful contributor to discussions (r = 0.86). Moreover, their pre-use expectations of LLM value align with post-use assessments (r = 0.61). These findings suggest that LLMs can enhance engagement and perceived learning when designed to promote equitable turn-taking and transparency across individual and shared spaces. The paper contributes an empirically grounded account of AI-mediated collaboration in a Global South higher-education context, with design implications for fairness-aware orchestration of human–AI teamwork.
\end{abstract}



\received{20 February 2007}
\received[revised]{12 March 2009}
\received[accepted]{5 June 2009}

\maketitle
\input{Sections/Introduction}

\input{Sections/RelatedWork}
\input{Sections/Methodology}
\input{Sections/Findings}

\input{Sections/Discussions}
\input{Sections/Conclusions}

\bibliographystyle{unsrt}
\bibliography{collaclass}

\end{document}

%% file: Sections/Introduction.tex
\section{Introduction}

Collaborative learning is a well-established pedagogical approach that fosters communication, problem-solving, and critical thinking among students \cite{Yang2023,Qureshi2023}.  At the same time, recent advances in AI and large language models (LLMs) promise to revolutionize learning by providing personalized, conversational support. For example, systems like OATutor and ChatScratch have demonstrated how LLMs can adapt content and feedback to individual learners’ needs \cite{pardos2023oatutor, chen2024chatscratch}.  However, existing AI-driven learning tools predominantly focus on one-on-one interactions, often neglecting the benefits of peer collaboration.  While some work explores AI to help teachers orchestrate transitions between individual and group work \cite{Yang2023PairUp}, there remains a lack of platforms that seamlessly blend human–human and human–AI collaboration in real time. This gap is especially consequential in contexts like Bangladesh. The higher-education system of this country has long faced a significant digital divide, with inequitable access to technology and online resources that limit educational equity \cite{Ahsan2023}.  At the same time, Bangladeshi students are beginning to embrace AI learning tools. Recent studies find that students view ChatGPT as efficient and useful for both in-class and out-of-class tasks \cite{Hasib2025}.  However, they also express concerns about responsible use and academic integrity.  These socio-technical factors – including resource constraints, large class sizes, and evolving attitudes toward AI – motivate the design of collaborative learning platforms that are equitable, accessible, and culturally appropriate.

\begin{figure*}[h!]
    \centering
    \fbox{\includegraphics[width=0.95\linewidth]{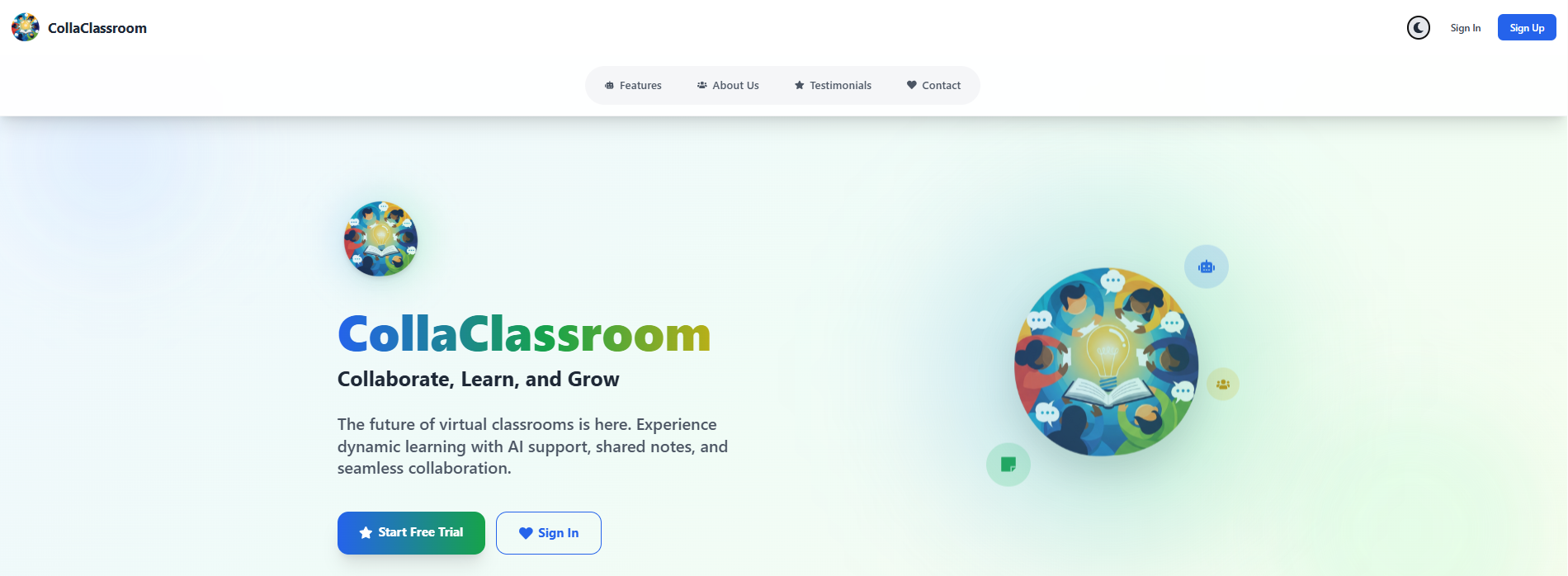}}\\[1em]
    \fbox{\includegraphics[width=0.95\linewidth]{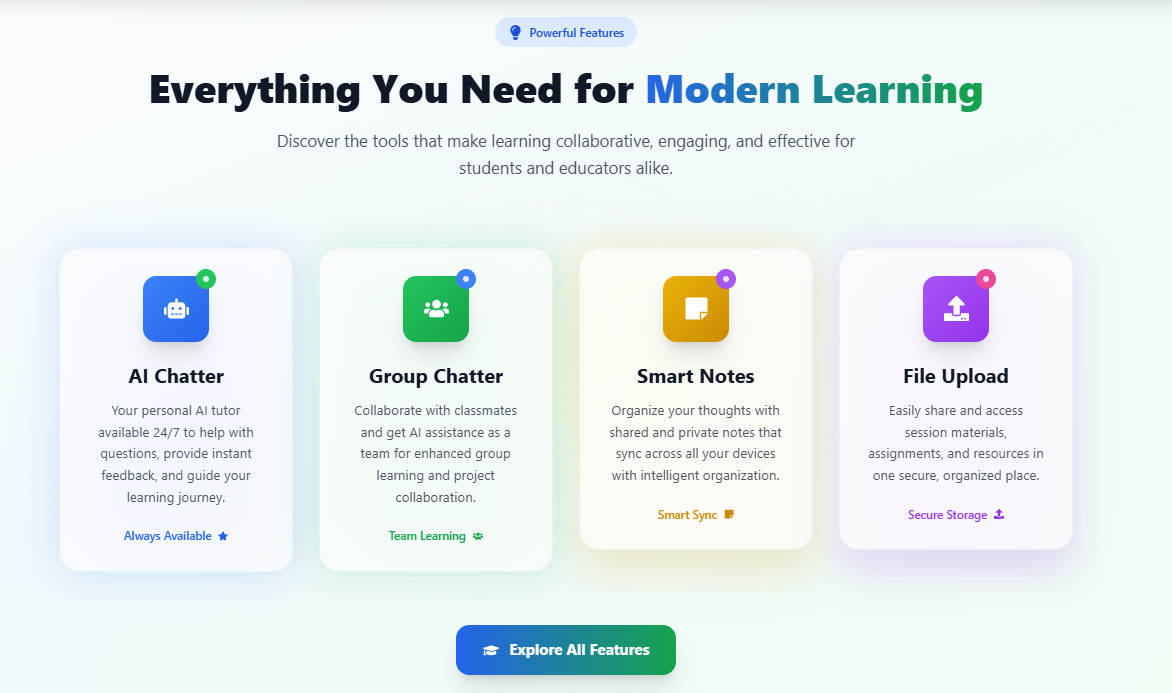}}
    \caption{Overview of \textit{CollaClassroom}, showing the main features}
    \label{fig:collaclassroom}
\end{figure*}

To address these challenges, we introduce \emph{CollaClassroom} (Figure \ref{fig:collaclassroom}), an AI-augmented collaborative learning platform tailored for Bangladeshi university students.  CollaClassroom provides a familiar virtual meeting interface (similar to Google Meet) with multiple synchronized panels. Two chat panels (\emph{My Chatter} and \emph{Group Chatter}) offer private and shared conversational spaces, and two note panels allow personal and group note-taking (Figure \ref{fig:colla_panels}).  Both chat panel can engage an LLM-based assistant grounded in the current session materials.  In \emph{My Chatter}, the LLM acts as a personal tutor, offering hints, clarifications, or autogenerated content to support individual work.  In \emph{Group Chatter}, the LLM functions as a co-author, helping moderate group discussion, suggest ideas, and integrate session content.  By giving each learner an equal voice in the shared environment and embedding AI support directly into both individual and group discourse, CollaClassroom operationalizes collaborative learning principles (e.g., participation equity and co-orchestration) at the interaction level.
We evaluate CollaClassroom through a user study with Bangladeshi university students.  Our investigation is guided by the following research questions:
\begin{itemize}
    \item \textbf{RQ1:} How do participants' prior experiences with collaborative learning and AI/LLM tools influence their receptivity to integrating AI into collaborative learning environments?
    \item \textbf{RQ2:} How does the integration of LLMs in collaborative learning impact participants' perceptions of equal participation, meaningful contribution, and overall learning effectiveness?
    \item \textbf{RQ3:} What are the key factors influencing system usability in AI-assisted collaborative learning, and how do they affect user satisfaction?
    \item \textbf{RQ4:} What correlations exist between students' perceptions of AI/LLM tool integration and their expectations regarding learning outcomes and system contribution?
\end{itemize}
This study makes the following contributions to the literature:
\begin{itemize}
    \item The design and implementation of \emph{CollaClassroom}, an AI-augmented collaborative learning environment that embeds LLM-based conversational agents in both individual and group contexts.
    \item An empirical evaluation with Bangladeshi university students, providing insights into user experiences with LLMs, collaboration perceptions, and system usability in a resource-constrained educational context.
    \item Design implications and guidelines for developing AI-supported educational technologies that promote personalization, accessibility, and equity in collaborative learning.
\end{itemize}

%% file: Sections/RelatedWork.tex
\section{Related work}
\subsection{Collaborative Learning}
Collaborative learning (CL) has evolved from distinct “collaborative” and “cooperative” traditions into a broad paradigm concerned with the joint construction of knowledge, often framed by social constructivism and enacted in computer-supported contexts \cite{yang2023historical,herrera2021collaborative}. While teachers routinely value CL, design practice still lags research prescriptions: activity structures, tool choices, and orchestration strategies vary widely across face-to-face, blended, and fully online settings \cite{pozzi2023collaborative}. Recent work on socio-spatial learning analytics further underscores that effective CL depends on how social interaction and physical/virtual space co-produce opportunities for participation, reflection, and teacher orchestration \cite{yan2023socio}. A substantial empirical base links CL to gains in higher-order outcomes. Studies report positive effects on critical thinking and its retention, as well as improvements in scientific literacy and affective engagement \cite{alharbi2022effect,warsah2021impact,dewi2021effect}. Effects are shaped by social factors—peer/teacher interaction, social presence, and even the productive use of social media—which mediate engagement and performance \cite{qureshi2023factors}. Pedagogical models that foreground the “4C” competencies (constructive, critical, creative, collaborative) and careful classroom design further enhance outcomes across cognitive, psychomotor, and affective domains \cite{supena2021influence}. In higher education, successful implementations emphasize supportive environments, phased group work, and authentic tasks, though instructors still report practical barriers and the need for replicable designs \cite{wicaksono2024impact,herrera2021collaborative}. Domain-specific reports (e.g., programming) echo these themes, showing improved participation and attainment when CL techniques are integrated into the curriculum \cite{sekhar2024collaborative}; at the group level, cohesion and synchronized interaction patterns correlate with higher performance \cite{zamecnik2024perceptions}.

Technology-mediated spaces broaden what “working together” can look like. Systematic reviews and platform proposals in immersive VR and the metaverse highlight shared problem spaces that can be engaging and reflective, yet they still struggle with usability and the reliable facilitation of social interaction \cite{paulsen2024designing,jovanovic2022vortex}. Structured methods like Jigsaw benefit from such spaces by enabling timely peer support, but orchestration remains a challenge \cite{chen2024enhancing}. Parallel efforts in human–AI co-orchestration show promise for dynamically transitioning learners between individual and collaborative modes, while revealing tensions around control and autonomy among teachers and students \cite{yang2023pair}. Against this backdrop, our work differs by centering orchestration \emph{within} the learners’ discourse: the \emph{Group Chatter} and \emph{My Chatter} of CollaClassroom provide simultaneous access to a shared and personal LLM, both grounded in the same session documents, enabling fluid movement between solo sensemaking and group dialogue without losing contextual continuity. This focus on a shared knowledge base accessible to multiple learners and the AI aims to operationalize the long-standing design aspirations of CL research—participation, equity of voice, and teacher/learner co-orchestration—directly in the conversational substrate of collaboration, rather than only at the level of tools or post-hoc analytics.

\begin{figure*}[h!]
    \centering    \fbox{\includegraphics[width=1.0\linewidth]{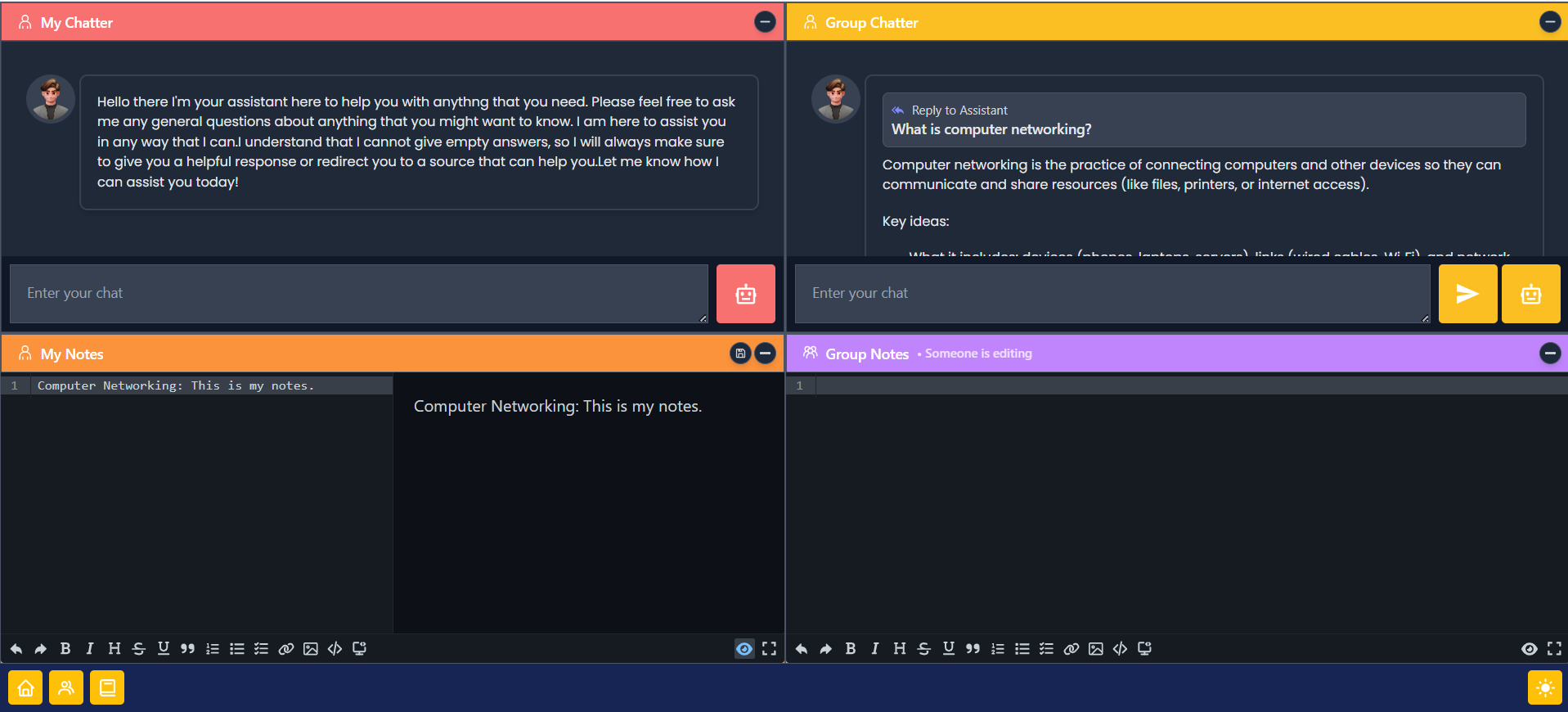}}
    \caption{User Interface of the four panels of the CollaClassroom}
    \label{fig:colla_panels}
\end{figure*}

\subsection{LLM in Education}
Large language models (LLMs) are reconfiguring the design and delivery of tutoring by lowering authoring costs and shifting from monologues to adaptive, dialogue-centered experiences. Systems now induce tutoring scripts from source materials and automate orchestration with multi-agent roles \cite{schmucker2023ruffle}, help instructors transform lecture videos into pedagogically grounded dialogues \cite{choi2024vivid}, and provide an open, extensible platform for replicable tutoring research \cite{pardos2023oatutor}. Together these lines of work move beyond static content toward scalable, conversational tutoring infrastructures.

Beyond infrastructure, LLMs are being embedded as learning companions across ages and domains. For early childhood mathematics, co-creative storytelling with an LLM can support vocabulary development on par with human partners, while shaping distinct engagement dynamics \cite{zhang2024mathemyths}. In K–12 computing, specialized assistants scaffold ideation, creative asset generation, and stepwise implementation to foster autonomous visual programming \cite{chen2024chatscratch}. In large university courses, assistants who withhold full solutions while offering conceptual guidance and annotated feedback suggest design principles for balancing help with productive struggle \cite{kazemitabaar2024codeaid}. These studies foreground interaction design choices—what to reveal, when to hint, and how to sustain agency—as central to LLM-mediated learning.

A complementary thread emphasizes structured scaffolds and methodological cautions. Domain ontologies paired with LLM support can accelerate understanding, analogical reasoning, and assessment in complex design curricula \cite{chen2024bidtrainer}; advances in language-based UI control broaden where and how learners can engage \cite{wang2023enabling}. While synthetic data generation promises rapid piloting yet raises validity concerns for educational research \cite{hamalainen2023evaluating}. Work on human–agent collaboration and peer-led critical pedagogies highlights the strategies learners adopt when “working with” conversational systems and the facilitation demands in online settings \cite{rapp2023collaborating,poon2023designing}. Against this backdrop, \emph{CollaClassroom} differs by centering LLMs within learner–learner discourse: it gives each student a personal assistant (\emph{My Chatter}) while also embedding a group assistant in a shared discussion space that is grounded in the same session documents. This dual-channel, shared-knowledge design aims to preserve individual agency and depth of inquiry without sacrificing the benefits of real-time collaborative sensemaking—an integration not directly addressed by prior LLM-in-education systems.

\subsection{AI adoption in education sector of Bangladesh}

The education sector of Bangladesh has seen initial efforts to integrate AI at the system level through institutional innovation and planning. For example, BRAC University has piloted AI-powered learning tools, but many public institutions still struggle with limited infrastructure and funding \cite{uddin2025artificial}. This disparity is driving calls for supportive frameworks and strategic planning in AI integration. Researchers have even proposed conceptual models to guide adoption; one such framework outlines key components like learner profiling, intelligent content delivery, and feedback mechanisms tailored to the local context \cite{rahmanai}. At the national level, policymakers are acknowledging the transformative potential of AI and advocating comprehensive strategies along with multi-stakeholder collaboration to ensure these technologies benefit all learners \cite{ahmed2022role, tarafdar2025artificial}. Collectively, these system-level efforts seek to modernize curricula and administration through AI, enabling more personalized learning and bridging resource gaps across the diverse educational landscape.

At the learner level, AI-powered tools are beginning to reshape student experiences and outcomes. Intelligent tutoring systems and adaptive learning platforms have shown potential to boost academic performance and tailor instruction to individual needs \cite{rahman2025impact, talukder2025impact}. Survey-based studies similarly report that AI-driven personalized learning correlates with greater student engagement and long-term educational success \cite{sultana2024does}. In practice, many students are already experimenting with AI applications: One recent campus survey has found that usage of a generative AI chatbot has risen from roughly 5\% of students in 2022 to nearly half by 2025 \cite{karmakerai}. Students' intent to use AI appears motivated by perceived learning benefits and peer influence more than by ease-of-use, and factors such as accessible content, institutional support, and AI awareness further shape their readiness to adopt new learning technologies \cite{jony2023intentions, akter2025readiness}.

Despite these encouraging developments, significant challenges and open questions remain. Uneven infrastructure and resource distribution pose major barriers: underfunded public institutions and rural schools often lack the connectivity and trained personnel needed to support AI initiatives at scale \cite{uddin2025artificial, rahman2025impact}. Educators and experts also highlight ethical and pedagogical concerns---data privacy, algorithmic bias, and the potential depersonalization of teaching are recurring issues that must be addressed for responsible AI implementation \cite{wafik2024academicians, tarafdar2025artificial}. There is caution that over-reliance on AI tools could undermine students' critical thinking or even encourage academic dishonesty \cite{talukder2025impact, karmakerai}. Moreover, AI adoption often mirrors existing digital divides: rural and disadvantaged learners have much lower access and usage rates compared to their urban peers, underscoring equity concerns \cite{karmakerai, rahman2025impact}. To navigate these issues, researchers advocate capacity-building programs, equitable infrastructure development, and contextually tailored policies to guide responsible AI integration \cite{rahman2025impact, wafik2024academicians}. In contrast to prior research---predominantly surveys, conceptual frameworks, and broad policy analyses---the CollaClassroom takes a more design-driven approach. It emphasizes collaborative human-AI interactions in real classroom settings, directly addressing pedagogical dynamics rather than treating AI adoption abstractly. By foregrounding interaction design and in-situ experimentation, CollaClassroom diverges from earlier frameworks and offers a practical pathway to harness AI's educational benefits while navigating these challenges.

\subsection{LLM Integration in Collaborative Learning}
Integrating Large Language Models (LLMs) into collaborative learning settings holds significant promise for transforming educational practices by enabling adaptive, intelligent agents that assist in group interactions. Several studies have explored the potential of LLMs in enhancing the collaborative learning experience, each addressing distinct aspects such as engagement, performance, and group dynamics. One notable approach incorporates LLMs into group discussions to facilitate collaborative problem-solving, offering real-time feedback, reflections, and guidance while maintaining a balance between personal autonomy and group collaboration \cite{tan2023leveraging,wei2024improving}. This integration not only enhances the students' ability to interact and negotiate knowledge but also improves cohesion within groups by supporting various learning styles and behavioral patterns \cite{hao2025student,zhang2025breaking}.
Other studies have focused on the use of LLM agents as moderators and active participants within the learning process. For instance, the introduction of peer agents—LLM-powered assistants—has proven effective in supporting collaborative interactions in children's learning environments by moderating discussions and guiding students through problem-solving tasks \cite{liu2024peergpt}. This dual role of LLMs, as both participants and moderators, enriches the social dimension of collaborative learning while addressing individual learners' needs. The results highlight the importance of timely intervention and the adaptability of LLM agents to the context of the group’s dialogue, which is a critical factor for maintaining engagement and enhancing learning outcomes \cite{cai2024advancing,sixu2024developing}.

Moreover, LLM-based systems have demonstrated substantial improvements in learning outcomes by providing adaptive support in problem-solving and reflection tasks. A study on LLM-powered virtual assistants showed that such models can guide students through complex tasks by providing contextually relevant assistance, enhancing collaborative learning performance, and boosting students' emotional satisfaction and willingness to learn \cite{wei2024improving}. Additionally, LLMs can serve as reflective tools in collaborative settings, prompting students to engage in deeper critical thinking and knowledge reorganization, further supporting the cognitive aspect of collaborative learning \cite{naik2025providing}. However, challenges remain, including the risks of cognitive overload and dependency, as excessive reliance on LLMs may reduce students' independence in learning processes \cite{zhang2025breaking}.
In contrast to these existing systems, CollaClassroom has a unique integration of LLM in both personal and group panels within a shared knowledge base enhances its effectiveness by allowing for continuous contextual relevance and fluid transitions between individual learning and group collaboration. This approach creates an ecosystem where both personal autonomy and group coherence are maintained, enabling students to benefit from tailored support while engaging in dynamic, real-time collaborative dialogue. This is a key difference from earlier LLM-based systems, which typically focus on one mode of interaction (either individual or group) or fail to integrate shared knowledge seamlessly into the learning process.

%% file: Sections/Methodology.tex
\section{Methodology}

\begin{figure*}[h!]
    \centering
    \fbox{\includegraphics[width=1.0\linewidth]{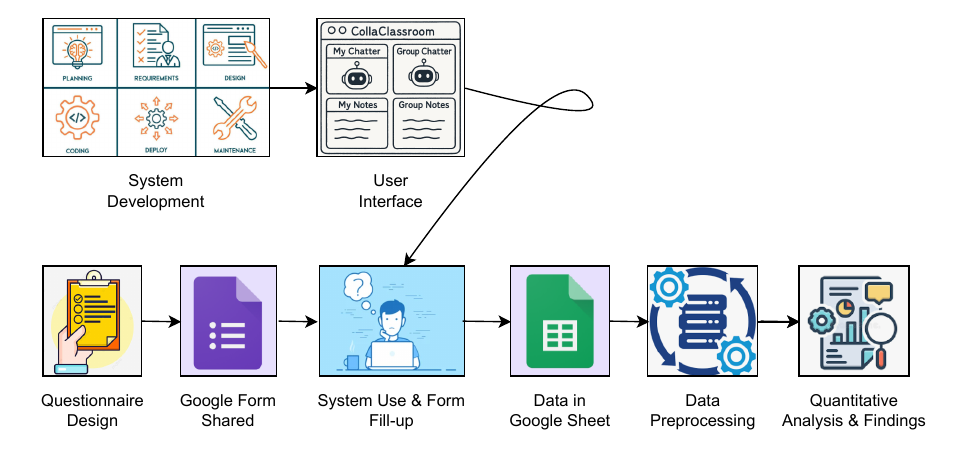}}
    \caption{Data Collection and Analysis Workflow of CollaClassroom}
    \label{fig:colla_method}
\end{figure*}

Our study does quantitative analysis, combining system deployment, survey responses and participant feedback to investigate the integration of large language models (LLMs) in collaborative learning environments. The methodology is designed to ensure ecological validity while aligning with established HCI research practices. Figure \ref{fig:colla_method} provides a detailed overview of the research methodology employed in this study.

\subsection{Participants}
We recruit twelve university students from Bangladeshi institutions, representing educational levels from undergraduate to doctoral studies, aged between 22 and 26 years. The majority of participants identify as male. This gender distribution aligns with prior findings indicating that gender does not significantly influence learning outcomes or user experience in educational technology \cite{seal2023effects, nizam2021derivation, kucuk2020students}. Therefore, the gender imbalance does not affect the validity of our study. The sample size follows established methodological standards in Human-Computer Interaction (HCI), where small-N exploratory studies are commonly used to evaluate early-stage systems. Prior work demonstrates that five to ten participants can uncover most usability issues \cite{faulkner2003beyond}, and subsequent research confirms that ten users can identify over 95\% of interface problems \cite{chandran2017sample}. Consistent with CHI community practices, we adopt a small sample size since each participant interacts with the system for 30–60 minutes, providing sufficient depth for qualitative insights and usability evaluation. Moreover, all participants are familiar with collaborative learning practices and have varying degrees of prior exposure to AI and LLM-based tools, allowing us to observe diverse interaction patterns in authentic learning contexts

\subsection{Participants Engagement}

After recruiting the participants, we initiate communication with each of them to explain the purpose of the study and the use of the CollaClassroom platform. Participants who agree to take part are listed and provided with a pre-study survey link to collect demographic information and baseline data on their prior experiences with AI and collaborative learning. Based on the responses, we organize participants into small groups of two to four members, ensuring shared academic backgrounds or common interests when possible to foster effective collaboration. For each group, a dedicated communication channel—either a Messenger or WhatsApp group—is created to facilitate coordination and discussion.
Within their groups, participants collectively decide on a discussion topic that will serve as the basis for their collaborative activity in the CollaClassroom system. Once the topic is agreed upon, participants register on the platform. One member from each group creates a new study session and shares the session invitation link with other group members through the group chat, allowing all members to join the same workspace.
Participants then engage with both the individual and group chat panels of CollaClassroom for approximately 30 to 60 minutes. After completing their collaborative session, they respond to the post-study survey, which collects feedback on usability, satisfaction, and perceived learning impact.

\subsection{Study Design}
The evaluation of CollaClassroom is structured around authentic learning activities to simulate real-world collaborative study sessions. Participants engage with the system in small groups, alternating between individual and group interactions with LLM-powered chat panels. The design emphasizes equal participation, co-orchestration, and equitable AI assistance across individual and shared spaces. To capture perceptions of usability, effectiveness, and group dynamics, we employ a pre–post survey design. The pre-survey establishes baseline information on participants’ demographics, prior experiences with AI and collaborative learning, and study preferences. The post-survey collects feedback on system usability, perceived usefulness, satisfaction, and learning outcomes.

\subsection{Survey Instrument}
The survey instrument is developed using Google Forms and included quantitative measures. Closed-ended questions assess collaborative learning frequency, AI/LLM tool usage, system usability, and perceived learning impact. Items are adapted from established frameworks such as the Technology Acceptance Model and usability heuristics. Open-ended questions invite participants to reflect on challenges, preferences, and perceptions of AI integration. The survey is piloted to ensure clarity and reliability before deployment.

\subsection{Data Collection and Procedure}
Participants complete the survey before and after using the system. During the study session, they interact with the dual-panel interface of the CollaClassroom, where both personal and group-level LLM assistants are available. The procedure lasted approximately one hour, with time allocated for orientation, collaborative activity, and debriefing. Responses are automatically logged through the survey platform and anonymized for analysis.

\subsection{Data Analysis}
Quantitative data are analyzed using descriptive statistics, chi-square tests, and pearson correlation analysis to examine relationships between prior experiences, perceptions of LLM contributions, and learning outcomes. Network graphs are generated to visualize correlations among satisfaction, perceived learning improvement, meaningful contribution, and equal participation.

\subsection{Ethical Considerations}
The study adheres to ethical research practices in HCI. Participants provide informed consent and are informed of their right to withdraw at any time. Data are anonymized and stored securely. The focus on Bangladeshi university students is carefully considered to ensure cultural appropriateness and sensitivity to resource-constrained educational contexts.

%% file: Sections/Findings.tex
\section{Findings}

We present our findings organized into four key areas: participant demographics, collaborative learning experiences, interactions with AI/LLM tools, and system usability evaluation. Our analysis draws on quantitative survey data from 12 participants and examines relationships between prior experience, perceived utility, and system satisfaction.

\subsection{Participant Demographics and Background}
Our study involves 12 university students, a sample size that aligns with established HCI research practices. A systematic review of 560 CHI papers finds sample sizes ranging from 1 to 916,000, with 12 being the most common value \cite{caine2016local}. Participants are primarily aged between 22 and 26 years (M=24.08, SD=1.16), with the majority identifying as male (83\%). Educational backgrounds vary from above college/diploma level (33\%) to PhD candidates (8\%), with half of the participants holding bachelor's degrees (50\%) and 17\% holding master's degrees. This diversity in academic experience provides varied perspectives on collaborative learning practices. Table~\ref{tab:demo-study-percentage} provides a complete demographic breakdown. 

\textbf{Prior Experience with Collaborative Learning and AI Tools.} The majority of participants report frequent engagement with collaborative learning, with 67\% indicating they ``often'' or ``always'' participate in collaborative study sessions. Similarly, 58\% of participants reported using AI/LLM tools ``often'' or ``always'' in their academic work, suggesting a sample with substantial prior exposure to both collaborative practices and AI-assisted learning tools. Notably, half of the participants (50\%) express a preference for studying collaboratively over working alone, while 25\% preferred solitary study and 25\% remained neutral.

\begin{table*}
\centering
\caption{Demographic information and study habits of participants (N=12).}
\label{tab:demo-study-percentage}
\begin{tabular}{@{}lcc@{}}
\toprule
\textbf{Variable} & \textbf{Category} & \textbf{Percentage} \\
\midrule
\multirow{5}{*}{Age (years)}
    & 22 & 17\% \\
    & 23 & 25\% \\
    & 24 & 25\% \\
    & 25 & 33\% \\
    & 26 & 8\% \\
\addlinespace[0.5em]
\multirow{2}{*}{Gender}
    & Male & 83\% \\
    & Female & 17\% \\
\addlinespace[0.5em]
\multirow{4}{*}{\parbox{3.5cm}{Educational\\Qualification}}
    & Above College/Diploma/Equivalent & 33\% \\
    & BSc/BA/Equivalent & 50\% \\
    & MSc/MA/Equivalent & 17\% \\
    & PhD/M.Phil/Postdoc/Equivalent & 8\% \\
\addlinespace[0.5em]
\multirow{4}{*}{\parbox{3.5cm}{Prior Experience with\\Collaborative Learning}}
    & Rarely & 8\% \\
    & Sometimes & 25\% \\
    & Often & 17\% \\
    & Always & 50\% \\
\addlinespace[0.5em]
\multirow{4}{*}{\parbox{3.5cm}{Prior Experience with\\AI/LLM Tools}}
    & Never & 17\% \\
    & Sometimes & 25\% \\
    & Often & 33\% \\
    & Always & 25\% \\
\addlinespace[0.5em]
\multirow{3}{*}{\parbox{3.5cm}{Study Preference:\\Collaborative vs. Alone}}
    & Mostly alone & 25\% \\
    & Neutral & 25\% \\
    & Mostly collaboratively & 50\% \\
\bottomrule
\end{tabular}
\end{table*}

\subsection{Collaborative Learning Experience and Effectiveness}

We examine the relationship between study preferences and perceived learning effectiveness through both frequency distributions and statistical analysis.

\textbf{Study Preferences and Perceived Impact.} When asked about their typical study patterns, participants are nearly evenly split: 50\% reported studying mostly alone, 41.67\% mostly collaboratively, and 8.33\% are neutral. Despite this distribution in practice, participants overwhelmingly recognize the value of collaborative learning. When comparing collaborative learning to learning alone, 58.33\% indicate that collaborative approaches lead to ``much better'' understanding, while 33.33\% find it ``somewhat better,'' and only 8.33\% perceived no difference (Table~\ref{tab:study-impact-separate}).

\textbf{Statistical Relationship Between Preference and Effectiveness.} To investigate whether study preferences correlate with perceived learning effectiveness, we conduct a chi-square test of independence. The analysis reveals no significant association between how often participants study collaboratively versus alone and their perceived effectiveness of collaborative learning ($\chi^2$(4) = 0.600, \textit{p} = 0.963). This null result suggests that recognition of collaborative learning's benefits exists independently of how frequently one engages in such practices—even those who primarily study alone acknowledge the effectiveness of collaboration (Table \ref{tab:chi-square-results}).

\begin{table*}[ht]
\centering
\caption{Distribution of study preferences and perceived impact of collaborative learning (N=12).}
\label{tab:study-impact-separate}
\begin{tabular}{@{}m{7cm}M{3.5cm}M{2.5cm}@{}}
\toprule
\textbf{Variable} & \textbf{Category} & \textbf{Percentage} \\
\midrule
\multirow{3}{7cm}{How often do you study collaboratively vs. alone?}
& Mostly alone & 50.00\% \\
& Neutral & 8.33\% \\
& Mostly collaboratively & 41.67\% \\
\midrule
\multirow{3}{7cm}{Compared to learning alone, how does collaborative learning impact your understanding?}
& About the same & 8.33\% \\
& Somewhat better & 33.33\% \\
& Much better & 58.33\% \\
\bottomrule
\end{tabular}
\end{table*}

\begin{table*}[ht]
\centering
\caption{Chi-square test results comparing study preference with learning effectiveness perception (N=12).}
\label{tab:chi-square-results}
\begin{tabular}{@{}lcccc@{}}
\toprule
\textbf{Variables Compared} & \textbf{$\chi^2$} & \textbf{\textit{p}-value} & \textbf{df} & \textbf{Result} \\
\midrule
\parbox{8cm}{Study preference (collaborative vs. alone) \\ 
\quad $\times$ \\ 
Perceived learning effectiveness} 
& 0.600 & 0.963 & 4 & n.s. \\
\bottomrule
\end{tabular}
\end{table*}

\subsection{Experience with AI/LLMs in Collaborative Contexts}

We explore participants' attitudes toward LLM integration and examine correlations between prior AI experience, perceived contributions, and learning outcomes.

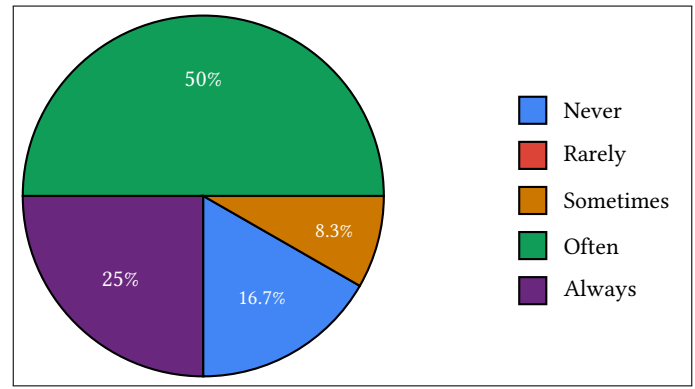
\begin{figure}[htbp]
\centering
\fbox{%
\begin{tikzpicture}[scale=1.2]
    \definecolor{never}{RGB}{66, 133, 244}
    \definecolor{rarely}{RGB}{219, 68, 55}
    \definecolor{sometimes}{RGB}{204, 120, 0}
    \definecolor{often}{RGB}{15, 157, 88}
    \definecolor{always}{RGB}{106, 39, 126}
    
    \fill[often] (0,0) -- (0:2) arc (0:180:2) -- cycle;
    \draw[black, thick] (0,0) -- (0:2) arc (0:180:2) -- cycle;
    
    \fill[always] (0,0) -- (180:2) arc (180:270:2) -- cycle;
    \draw[black, thick] (0,0) -- (180:2) arc (180:270:2) -- cycle;
    
    \fill[never] (0,0) -- (270:2) arc (270:330.12:2) -- cycle;
    \draw[black, thick] (0,0) -- (270:2) arc (270:330.12:2) -- cycle;
    
    \fill[sometimes] (0,0) -- (330.12:2) arc (330.12:360:2) -- cycle;
    \draw[black, thick] (0,0) -- (330.12:2) arc (330.12:360:2) -- cycle;
    
    \node[text=white] at (90:1.3) {50\%};
    \node[text=white] at (225:1.3) {25\%};
    \node[text=white] at (300:1.3) {\small16.7\%};
    \node[text=white] at (345:1.5) {\small8.3\%};
    
    \begin{scope}[shift={(3.5,0.8)}]
        \fill[never] (0,0) rectangle (0.3,0.3);
        \draw[black, thick] (0,0) rectangle (0.3,0.3);
        \node[anchor=west] at (0.4,0.15) {Never};
        
        \fill[rarely] (0,-0.5) rectangle (0.3,-0.2);
        \draw[black, thick] (0,-0.5) rectangle (0.3,-0.2);
        \node[anchor=west] at (0.4,-0.35) {Rarely};
        
        \fill[sometimes] (0,-1.0) rectangle (0.3,-0.7);
        \draw[black, thick] (0,-1.0) rectangle (0.3,-0.7);
        \node[anchor=west] at (0.4,-0.85) {Sometimes};
        
        \fill[often] (0,-1.5) rectangle (0.3,-1.2);
        \draw[black, thick] (0,-1.5) rectangle (0.3,-1.2);
        \node[anchor=west] at (0.4,-1.35) {Often};
        
        \fill[always] (0,-2.0) rectangle (0.3,-1.7);
        \draw[black, thick] (0,-2.0) rectangle (0.3,-1.7);
        \node[anchor=west] at (0.4,-1.85) {Always};
    \end{scope}
\end{tikzpicture}
}
\caption{Prior experience with AI/LLM tools in study/work (N=12)}
\label{fig:ai_experience}
\end{figure}

\textbf{Prior AI/LLM Experience Patterns.} Analysis of participants' prior experience with AI/LLM tools reveals a distribution weighted toward frequent use: 50\% report using such tools ``often,'' 25\% ``always,'' 8.3\% ``sometimes,'' and 16.7\% ``never'' (Figure~\ref{fig:ai_experience}). This distribution indicates that three-quarters of our sample have substantial prior experience with AI tools, providing them with a foundation for evaluating LLM integration in collaborative settings.

\begin{figure*}[h!]
\centering
\begin{tikzpicture}
\begin{scope}[local bounding box=chart]
\begin{axis}[
    ybar,
    symbolic x coords={Strongly agree, Agree, Neutral, Disagree, Strongly disagree},
    xtick=data,
    ylabel={Percentage (\%)},
    title={Opinion on Integrating LLM into Study Sessions},
    ymin=0, ymax=65,
    bar width=20pt,
    nodes near coords,
    nodes near coords align={vertical},
    enlarge x limits=0.4,
    grid=major,
    x tick label style={rotate=45, anchor=east},
    width=0.6\textwidth,
    ymajorgrids=true,
    axis lines=box,
    axis line style={thick, black},
    every outer x axis line/.append style={thick, black},
    every outer y axis line/.append style={thick, black},
]
\addplot[
    fill=blue!50,
    draw=black,
    thick
] coordinates {
    (Strongly agree, 33.33) 
    (Agree, 58.33) 
    (Neutral, 8.33) 
    (Disagree, 0) 
    (Strongly disagree, 0)
};
\end{axis}
\end{scope}
\draw[thick] ($ (chart.south west)+(-0.4,-0.1) $) rectangle ($ (chart.north east)+(1.1,0.5) $);
\end{tikzpicture}
\caption{Would integrating an LLM into your study sessions help? (N=12)}
\label{fig:llm-integration}
\end{figure*}
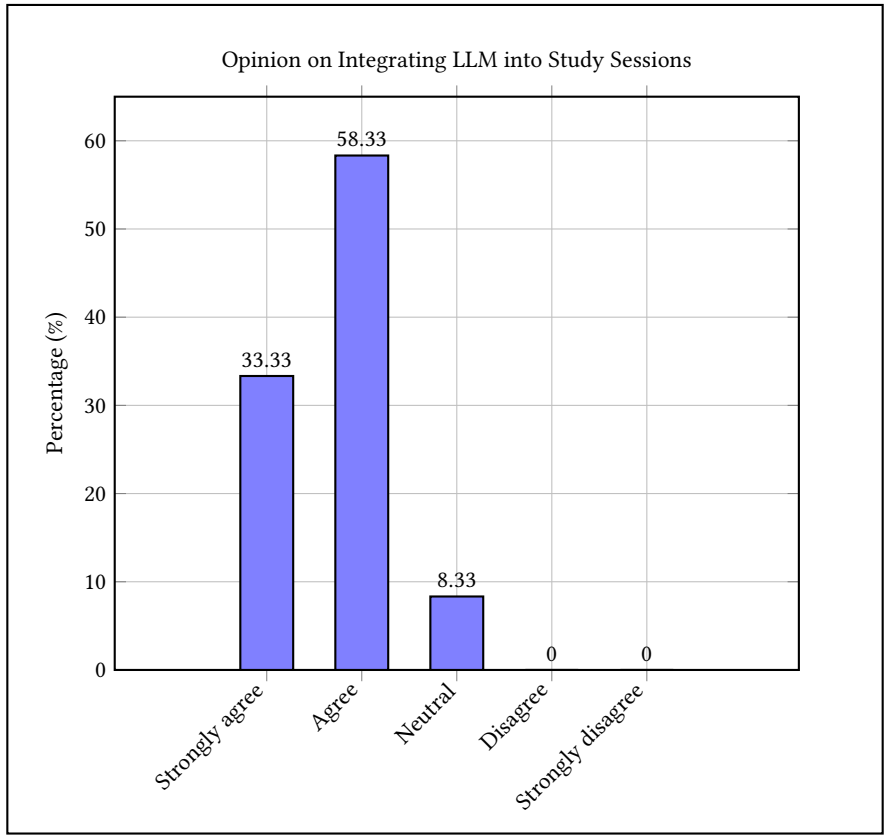

\textbf{Receptivity to LLM Integration.} Participants demonstrate strong receptivity to integrating LLMs into study sessions, with 91.66\% responding positively (33.33\% ``strongly agree,'' 58.33\% ``agree'') to the question of whether such integration would be helpful (Figure~\ref{fig:llm-integration}). Only 8.33\% remain neutral, and no participants disagree. This nearly unanimous positive sentiment suggests openness to AI-augmented collaborative learning among students with varied levels of prior AI exposure.

\textbf{Correlational Analysis of Key Factors.} To understand relationships between experience, expectations, and outcomes, we conduct Pearson correlation analyses on several key variables (Table~\ref{tab:correlation_analysis}). First, we examine the relationship between prior experience with AI/LLM tools and prior experience with collaborative learning, finding a moderate positive correlation (\textit{r} = 0.4117). This suggests that students who frequently engage in collaborative learning also tend to have more experience with AI tools, possibly indicating a general openness to diverse learning modalities.
Second, we find a moderate to strong positive correlation (\textit{r} = 0.6098) between participants' belief that LLM integration would help their study sessions (measured before the study) and their perception that the LLM contributed meaningfully to group discussion (measured after using the system). This consistency between expectation and experience suggests that the system meets or aligns with participants' anticipated benefits. Most notably, we observe a very strong positive correlation (\textit{r} = 0.8554) between participants' perception that the LLM supported equal participation and their assessment of the LLM's meaningful contribution to discussion. This strong association indicates that participants who perceive the LLM as supporting equitable participation also tend to rate its contribution to discussion as more meaningful.

\begin{table*}[htbp]
\centering
\caption{Pearson Correlation Analysis of LLM Integration and Collaborative Learning Factors}
\label{tab:correlation_analysis}
\begin{threeparttable}
\begin{tabular}{@{}p{5cm}p{5cm}cc@{}}
\toprule
\textbf{Variable 1} & \textbf{Variable 2} & \textbf{Correlation} & \textbf{Strength} \\
\midrule
Prior experience with AI/LLM tools & Prior experience with collaborative learning & 0.4117 & Moderate positive \\
\addlinespace
Do you think integrating an LLM into your study sessions would help? & The LLM contributed meaningfully to the group discussion & 0.6098 & Moderate to strong positive \\
\addlinespace
I felt that the LLM supported equal participation within the group & The LLM contributed meaningfully to the group discussion & 0.8554 & Very strong positive \\
\bottomrule
\end{tabular}
\end{threeparttable}
\end{table*}

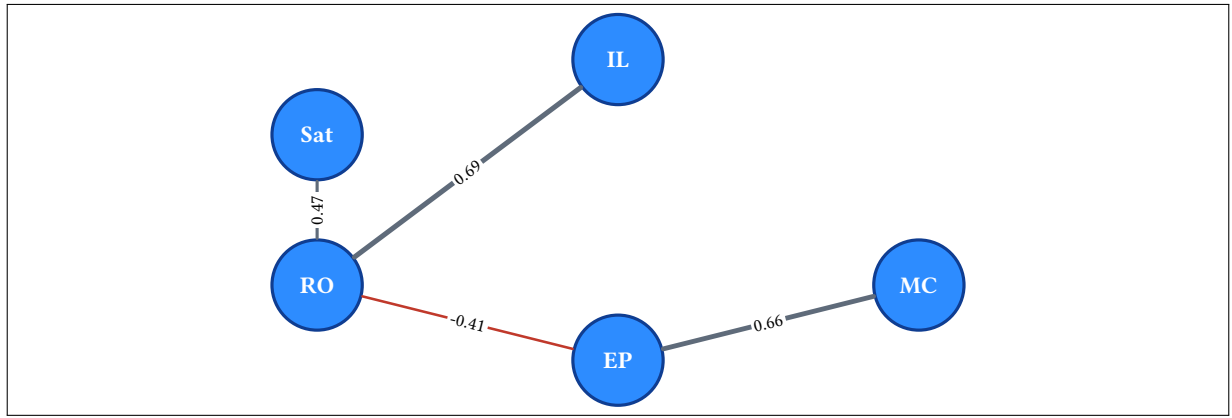
\begin{figure*}[t]
  \centering
  \fbox{%
    \begin{minipage}{0.9\linewidth}  
      \centering
      \begin{tikzpicture}
        \node[corrnode] (RO)  at (0,0)   {RO};
        \node[corrnode] (Sat) at (0,2)   {Sat};
        \node[corrnode] (IL)  at (4,3)   {IL};
        \node[corrnode] (EP)  at (4,-1)  {EP};
        \node[corrnode] (MC)  at (8,0)   {MC};

        \corrEdge{RO}{Sat}{0.47}{0.47}{corrpos}
        \corrEdge{RO}{IL}{0.69}{0.69}{corrpos}
        \corrEdge{RO}{EP}{-0.41}{-0.41}{corrneg}
        \corrEdge{EP}{MC}{0.66}{0.66}{corrpos}
      \end{tikzpicture}
    \end{minipage}%
  }

  \vspace{0.5\baselineskip}
  \caption{This network graph visualizes the relationships between key survey questions on LLM integration in collaborative learning, with the following mappings: Sat (`I am satisfied with it'), RO (`I would recommend it to a friend/classmate'), MC (`The LLM contributed meaningfully to the group discussion'), IL (`Working with both human peers and the LLM improved my learning experience'), and EP (`I felt that the LLM supported equal participation within the group'). The edges represent correlation strengths, with green indicating strong positive correlations (greater than 0.6), blue for moderate correlations (0.4-0.6), and gray for weak correlations (less than or equal to 0.4).}
  \label{fig:net_graph}
\end{figure*}

\textbf{Network Analysis of Collaborative Learning Factors.} 
To visualize the interconnections among participants' perceptions, a network graph is constructed examining relationships between five key constructs: \textit{satisfaction}, \textit{recommendation to others}, \textit{meaningful contribution}, \textit{improved learning}, and \textit{equal participation support} (Figure~\ref{fig:net_graph}). 
The graph highlights two strong positive correlations (green edges, $r > 0.6$): between \textit{improved learning} and \textit{recommendation to others} ($r = 0.69$), and between \textit{equal participation} and \textit{meaningful contribution} ($r = 0.66$). 
These strong links suggest that perceiving equitable participation and meaningful contributions are closely tied to both enhanced learning and the likelihood of recommending the system to others. 
A moderate positive relationship (blue edge, $r = 0.47$) appears between \textit{satisfaction} and \textit{recommendation to others}, indicating that general satisfaction contributes to endorsement but is not its sole driver.

\subsection{System Usability Evaluation}

We assess the system across six dimensions adapted from established usability frameworks: terminology and interface clarity, learnability, system capabilities, perceived usefulness, satisfaction, and cognitive load.

\textbf{Terminology and Interface.} Participants generally find the system's terminology consistent and appropriate, with 75\% agreeing that ``use of terms throughout the system'' is clear and only 25\% remaining neutral. Regarding error messages, half of the participants (50\%) report encountering no errors at all, while 25\% encounter them slightly and 25\% moderately (Table~\ref{tab:terminology-interface}). 

\begin{table*}[ht]
\centering
\caption{System Evaluation (Usability \& Experience) - Terminology \& Interface}
\label{tab:terminology-interface}
\begin{tabular}{@{}m{8cm}M{3cm}M{2.5cm}@{}}
\toprule
\textbf{Variables} & \textbf{Category} & \textbf{Frequency (\%)} \\
\midrule
\multirow{2}{8cm}{Use of terms throughout the system} & Agree & 75.00 \\
& Neutral & 25.00 \\
\midrule
\multirow{3}{8cm}{Error messages} & Not at all & 50.00 \\
& Slightly & 25.00 \\
& Moderately & 25.00 \\
\bottomrule
\end{tabular}
\end{table*}

\textbf{Learnability.} The system demonstrates strong learnability characteristics. Two-thirds of participants (66.67\%) find learning to operate the system ``easy,'' with the remaining third (33.33\%) neutral—notably, no participants rate it as difficult. Similarly, 83.33\% agree that they ``easily remember how to use it,'' with 16.67\% neutral (Table~\ref{tab:learnability}). These results indicate that the system presents a low learning curve, an important factor for adoption in educational contexts where time constraints are significant.

\begin{table*}[ht]
\centering
\caption{System Evaluation (Usability \& Experience) - Learnability}
\label{tab:learnability}
\begin{tabular}{@{}m{8cm}M{3cm}M{2.5cm}@{}}
\toprule
\textbf{Variables} & \textbf{Category} & \textbf{Frequency (\%)} \\
\midrule
\multirow{2}{8cm}{Learning to operate the system} & Easy & 66.67 \\
& Neutral & 33.33 \\
\midrule
\multirow{2}{8cm}{I easily remember how to use it} & Agree & 83.33 \\
& Neutral & 16.67 \\
\bottomrule
\end{tabular}
\end{table*}

\textbf{System Capabilities.} Participants rate system performance positively across both speed and reliability dimensions. For system speed, 66.67\% rate it as ``fast'' and 33.33\% as ``moderate,'' with no reports of slow performance. System reliability receives even stronger ratings, with 83.33\% finding it ``reliable'' and 16.67\% ``somewhat reliable'' (Table~\ref{tab:capabilities}). The absence of ratings below ``somewhat reliable'' suggests consistent technical performance throughout the study.

\begin{table*}[ht]
\centering
\caption{System Evaluation (Usability \& Experience) - System Capabilities}
\label{tab:capabilities}
\begin{tabular}{@{}m{8cm}M{3cm}M{2.5cm}@{}}
\toprule
\textbf{Variables} & \textbf{Category} & \textbf{Frequency (\%)} \\
\midrule
\multirow{2}{8cm}{System speed} & Fast & 66.67 \\
& Moderate & 33.33 \\
\midrule
\multirow{2}{8cm}{System reliability} & Reliable & 83.33 \\
& Somewhat Reliable & 16.67 \\
\bottomrule
\end{tabular}
\end{table*}

\textbf{Perceived Usefulness.} Drawing from the Technology Acceptance Model framework, we assess perceived usefulness through two key dimensions. For the statement ``using the system would make it easier to do my work/study,'' 75\% agree, 16.67\% remained neutral, and 8.33\% strongly agree. Regarding whether the system ``does everything I expect it to do,'' 58.33\% agree, 33.33\% are neutral, and 8.33\% strongly agree (Table~\ref{tab:usefulness}). While responses are predominantly positive, the higher neutral percentage on the second item suggests some participants may have had expectations that are not fully meet, pointing to opportunities for feature enhancement.

\begin{table*}[ht]
\centering
\caption{System Evaluation (Usability \& Experience) - Perceived Usefulness (PUEU)}
\label{tab:usefulness}
\begin{tabular}{@{}m{8cm}M{3cm}M{2.5cm}@{}}
\toprule
\textbf{Variables} & \textbf{Category} & \textbf{Frequency (\%)} \\
\midrule
\multirow{3}{8cm}{Using the system would make it easier to do my work/study} & Agree & 75.00 \\
& Neutral & 16.67 \\
& Strongly Agree & 8.33 \\
\midrule
\multirow{3}{8cm}{It does everything I expect it to do} & Agree & 58.33 \\
& Neutral & 33.33 \\
& Strongly Agree & 8.33 \\
\bottomrule
\end{tabular}
\end{table*}

\textbf{User Satisfaction.} Overall satisfaction ratings are positive, with 66.67\% agreeing with the statement ``I am satisfied with it'' and 33.33\% neutral. Willingness to recommend the system is slightly higher, with 75\% agreeing they ``will recommend it to a friend/classmate'' and 25\% neutral (Table~\ref{tab:satisfaction}). The absence of disagreement on both measures, combined with the higher recommendation rate, suggests that participants see value in the system for peer contexts even when their own satisfaction is moderate.

\begin{table*}[ht]
\centering
\caption{System Evaluation (Usability \& Experience) - Satisfaction (USE)}
\label{tab:satisfaction}
\begin{tabular}{@{}m{8cm}M{3cm}M{2.5cm}@{}}
\toprule
\textbf{Variables} & \textbf{Category} & \textbf{Frequency (\%)} \\
\midrule
\multirow{2}{8cm}{I am satisfied with it} & Agree & 66.67 \\
& Neutral & 33.33 \\
\midrule
\multirow{2}{8cm}{I would recommend it to a friend/classmate} & Agree & 75.00 \\
& Neutral & 25.00 \\
\bottomrule
\end{tabular}
\end{table*}

\textbf{Cognitive Load.} We assess cognitive load through the effort required and the frustration experienced. Half of participants (50\%) report ``moderate'' effort needed to use the system effectively, while 33.33\% report ``low'' effort and 16.67\% ``high'' effort. Importantly, frustration levels are minimal: 83.33\% feel ``not at all'' frustrated while using the system, and only 16.67\% feel ``slightly'' frustrated, with no reports of moderate or high frustration (Table~\ref{tab:cognitive-load}). The low frustration despite some reported effort suggests that while the system may require cognitive engagement, it does not create negative emotional responses—an important distinction for learning contexts where productive struggle differs from frustrating obstacles.

\begin{table*}[ht]
\centering
\caption{System Evaluation (Cognitive Load)}
\label{tab:cognitive-load}
\begin{tabular}{@{}m{8cm}M{3cm}M{2.5cm}@{}}
\toprule
\textbf{Variables} & \textbf{Category} & \textbf{Frequency (\%)} \\
\midrule
\multirow{3}{8cm}{How much effort did it take to use the system effectively?} & Low & 33.33 \\
& Moderate & 50.00 \\
& High & 16.67 \\
\midrule
\multirow{2}{8cm}{How frustrated did you feel while using the system?} & Not at all & 83.33 \\
& Slightly & 16.67 \\
\bottomrule
\end{tabular}
\end{table*}

Our findings reveal a sample of university students with substantial prior experience in both collaborative learning and AI tool usage who demonstrate strong receptivity to LLM integration in study sessions. While study preferences vary, participants consistently recognize the effectiveness of collaborative learning. Notably, perceptions of equal participation emerge as strongly correlated with assessments of the meaningful contribution of the LLM, suggesting that supporting equitable engagement is crucial for successful AI integration in collaborative contexts. System usability evaluations are predominantly positive across all dimensions, with particularly strong ratings for learnability and reliability, though some participants report moderate cognitive effort requirements. The low frustration levels despite this effort suggest that the system's cognitive demands align with productive learning engagement rather than unnecessary complexity.

%% file: Sections/Discussions.tex
\begin{table*}[h!]
\centering
\small
\caption{Comparison of CollaClassroom with related LLM-based educational systems.}
\label{tab:comparison}
\begin{tabular}{|p{2.0cm}|p{2.0cm}|p{2.0cm}|p{3.5cm}|p{3.5cm}|}
\hline
\textbf{Study} & \textbf{Audience/ Context} & \textbf{LLM Role / Design} & \textbf{Key features} & \textbf{Contrast with CollaClassroom} \\ \hline
\textbf{Ruffle \& Riley} \cite{schmucker2023ruffle} & Adult learners in online tutoring & Two LLM agents (student \& professor) automatically induce a tutoring script from lesson text & Automatic script induction; LLMs orchestrate a free-form conversation; improves perceived understanding though no significant learning advantage & Single-learner scenario with no group chat; lacks dual-channel assistance; does not integrate a shared knowledge base like CollaClassroom \\ \hline
\textbf{VIVID} \cite{choi2024vivid} & Instructors authoring lecture dialogues & LLM collaborates with teachers to transform monologue lectures into pedagogical dialogues & Applies design guidelines to convert lecture scripts into vicarious dialogues; supports instructor-led editing and evaluation & Focuses on authoring offline dialogues rather than real-time collaboration; no learner-facing multi-user interaction; CollaClassroom provides synchronous group and personal channels \\ \hline
\textbf{ChatScratch} \cite{chen2024chatscratch} & Children (6–12 yrs) learning Scratch & Scratch-specific LLM with voice interface & Structured storyboards; AI-powered asset creation with Stable Diffusion and digital drawing; step-by-step coding guidance; improves project quality & Supports autonomous individual programming, not group collaboration; no shared context across users; CollaClassroom targets group sensemaking and co-orchestration \\ \hline
\textbf{Mathemyths} \cite{zhang2024mathemyths} & Children (4–8 yrs) & LLM as co-creative story partner & Co-creates stories that embed mathematical terms; uses prompt-engineering; yields comparable language learning to human partner & Emphasises one-on-one storytelling; no multi-learner coordination; CollaClassroom integrates personal and group LLMs anchored on shared documents \\ \hline
\textbf{PeerGPT} \cite{liu2024peergpt} & Children’s collaborative workshops & LLM-based peer agent as moderator or participant & Moderates or participates in group discussions; fosters creative thinking but sometimes lacks timely feedback & Explores moderation but does not combine personal and group channels or share document context; CollaClassroom uses dual LLMs grounded in the same session documents \\ \hline
\textbf{CodeAid} \cite{kazemitabaar2024codeaid} & University programming course & LLM-powered programming assistant (GPT-3.5) & Helps students without revealing code; features include general questions, code explanations, interactive pseudo-code, and annotated errors; deployed in a large class & Designed for individual help and academic integrity; not collaborative; CollaClassroom supports real-time group work and fluid transitions between individual and group interactions \\ \hline
\textbf{CollaClassroom} & University Students (Bangladesh) & LLMs as personal \& group assistants grounded on shared session documents & Four panels—Group Chatter, My Chatter, My Note, Group Note; supports fluid transitions between solo and group sensemaking; emphasises participation, equity of voice and teacher/learner co-orchestration & Integrates group and personal LLMs within a shared knowledge base; enables synchronous collaboration and individual reflection; evaluated in Bangladeshi classrooms \\ \hline
\end{tabular}
\end{table*}

\section{Discussion}


The findings from this study provide valuable insights into how prior experiences with collaborative learning and AI/LLM tools, as well as perceptions of system usability, impact the integration of AI into collaborative learning environments. As summarized in Table~\ref{tab:comparison}, CollaClassroom differs from prior LLM-based learning systems by embedding dual conversational agents—one personal and one group-facing—within a shared workspace. This dual-channel design supports both individualized reflection and synchronous group sensemaking, extending previous single-user or instructor-mediated paradigms. In this section, we discuss the key findings in relation to the research questions posed in this study.

\subsection{Prior Experience and Receptivity to LLM Integration (RQ1)}
Participants report strong receptivity to LLM augmentation in study sessions (Figure~\ref{fig:llm-integration}). This aligns with the sample’s substantial exposure to both collaborative learning and AI tools (Table~\ref{tab:demo-study-percentage}). The observed moderate association between prior AI/LLM use and prior collaborative learning (\emph{r} = 0.4117; Table~\ref{tab:correlation_analysis}) indicates that students familiar with one modality also tend to engage with the other. While not evidencing causality, this pattern suggests overlapping dispositions toward technology-supported collaboration and may help explain the broadly positive attitudes toward integrating LLMs into group work.

\subsection{Perceived Effects on Participation and Contribution (RQ2)}
Participants rate LLM support as aligning with collaborative effectiveness: perceived helpfulness prior to use corresponds with post-use judgments that the LLM contributed meaningfully to discussion (\emph{r} = 0.6098). Notably, perceptions of equal participation show a very strong association with perceived meaningful contribution (\emph{r} = 0.8554). Taken together, these patterns indicate that evaluations of AI support in group settings cohere around two linked facets: equitable participation and visible, discussion-level contribution. Consistent with this, the correlation network (Figure~\ref{fig:net_graph}) highlights strong ties between improved learning, recommendation intent, and judgments of equal participation and meaningful contribution. We do not infer directionality; rather, the results point to a tightly coupled perception space in which equity and contribution co-occur with positive learning appraisals.

\subsection{Usability, Satisfaction, and Cognitive Effort (RQ3)}
System usability is rated positively across terminology, learnability, speed, and reliability (Tables~\ref{tab:terminology-interface}--\ref{tab:capabilities}). Most participants describe the system as easy to learn and easy to remember (Table~\ref{tab:learnability}), and reliability ratings are high. Satisfaction and willingness to recommend are similarly positive (Table~\ref{tab:satisfaction}). Reports of moderate effort by some participants (Table~\ref{tab:cognitive-load}) coexist with low frustration, suggesting productive engagement rather than detrimental overload. Neutral responses on ``does everything I expect'' (Table~\ref{tab:usefulness}) point to opportunity areas for feature breadth or depth while maintaining a favorable baseline of usability.

\subsection{Expectations and Experience Alignment (RQ4)}
The moderate-to-strong association between pre-study expectations of benefit and post-study perceptions of meaningful contribution (\emph{r} = 0.6098) indicates expectation--experience alignment for many participants. In parallel, the very strong association between equal participation and meaningful contribution (\emph{r} = 0.8554) underscores that judgments of LLM value in group settings are closely linked with perceived fairness of participation. While these are correlational results, they highlight evaluation criteria that appear central to how learners appraise AI in collaborative contexts.

\subsection{Design Implications}
Based on the findings, we identify several directions for designing future AI-supported collaborative learning systems.

\noindent\textbf{Encourage balanced participation.}
Because equal participation is strongly linked to how useful the LLM feels, future systems include features that help everyone contribute—such as reminders to involve quieter members or short summaries of who participates most.

\noindent\textbf{Make the AI’s role visible.}
Students view the LLM more positively when its ideas are clear and easy to understand. Systems highlight when and why the AI suggests something, helping learners see it as a transparent and fair partner in group work.

\noindent\textbf{Keep the system simple to learn.}
Participants find CollaClassroom easy to learn and remember. Future tools maintain this simplicity by using consistent language and predictable actions, even as new features are added.

\noindent\textbf{Match support to user expectations.}
Some students expect more features than the system provides. Allowing customization—such as switching between “idea support” and “critique” modes—helps align AI assistance with different study needs.

Together, these implications highlight that fairness, clarity, and simplicity are essential for building trust and engagement in AI-assisted collaborative learning.

\subsection{Limitations and Future Work}
This study involves a small, convenience sample (N=12) and focuses on short, single-session interactions; results may not generalize to larger cohorts, extended deployments, or other educational settings. Findings are based on self-report measures and correlational analyses; we avoid causal claims. Future work could incorporate complementary behavioral or conversational analytics, controlled manipulations of participation-support features, and longer-term classroom studies to assess durability of effects and learning transfer.

%% file: Sections/Conclusions.tex
\section{Conclusion}
This study explore university students' experiences with collaborative learning and the integration of AI/LLM tools, focusing on their prior experiences, perceptions of learning effectiveness, system usability, and equitable participation in group discussions. Our findings indicate that students with prior exposure to collaborative learning and AI tools are highly receptive to AI-assisted collaborative learning, recognizing its potential to enhance understanding and engagement.  The integration of LLMs is found to positively influence learning outcomes, particularly when the system supports equal participation and contributes meaningfully to discussions. Strong correlations between perceived equitable participation and meaningful contribution highlight the importance of designing AI systems that promote fairness and inclusivity in collaborative learning environments. System usability evaluations reveal that participants find the LLM system easy to learn, reliable, and cognitively manageable, with minimal frustration reported. These findings suggest that well-designed AI tools can integrate seamlessly into educational contexts, providing both functional support and an enhanced collaborative experience.  Overall, this research demonstrates the promise of AI/LLM tools in supporting collaborative learning among university students. By fostering equitable participation, meaningful contributions, and positive learning outcomes, AI systems can serve as effective partners in educational settings. Future work should focus on scaling such systems for larger groups, refining features to better meet user expectations, and exploring the impact of AI-assisted collaboration across diverse learning contexts and disciplines.